\documentclass[a4paper]{jpconf}
\usepackage{graphicx}

\begin{document}

\title{Indications for the onset of deconfinement
  in nucleus nucleus collisions}

\begin{center}
D. Flierl\footnote[1]{presented at ICHEP 2004 conference} for the NA49 collaboration
\end{center}

\author{
C.~Alt$^{9}$, T.~Anticic$^{21}$, B.~Baatar$^{8}$,D.~Barna$^{4}$,
J.~Bartke$^{6}$, 
L.~Betev$^{9,10}$, H.~Bia{\l}\-kowska$^{19}$, A.~Billmeier$^{9}$,
C.~Blume$^{9}$,  B.~Boimska$^{19}$
% ...
}

\begin{center}
(NA49 Collaboration)
\end{center}

\address{
% ...
$^{4}$KFKI Research Institute for Particle and Nuclear Physics, Budapest, Hungary.\\
% ...
$^{6}$Institute of Nuclear Physics, Cracow, Poland.\\
% ...
$^{8}$Joint Institute for Nuclear Research, Dubna, Russia.\\
$^{9}$Fachbereich Physik der Universit\"{a}t, Frankfurt, Germany.\\
$^{10}$CERN, Geneva, Switzerland.\\
% ...
$^{19}$Institute for Nuclear Studies, Warsaw, Poland.\\
% ...
$^{21}$Rudjer Boskovic Institute, Zagreb, Croatia.\\
}
\ead{flierl@ikf.uni-frankfurt.de}

\begin{abstract}
The hadronic final state of central Pb+Pb collisions at 20, 30, 40, 80, and 158 AGeV has been measured by the CERN NA49 collaboration. 
The mean transverse mass of pions and kaons at midrapidity stays nearly constant in this energy range, whereas at 
lower energies, at the AGS, a steep increase with beam energy was measured.
Compared to p+p collisions as well as to model calculations, anomalies in the energy dependence of pion and kaon production at lower SPS energies are observed.
These findings can be explained, assuming that the energy density reached in central A+A collisions at lower SPS energies is sufficient
to transform the hot and dense nuclear matter into a deconfined phase.
\end{abstract}

\section{Introduction}
At top CERN SPS and at RHIC energies the initial energy density reached in heavy ion collisions is apparently large enough to create a deconfined phase 
at the early stage of the collision: the Quark Gluon Plasma\cite{Heinz:2000bk}\raisebox{1ex}{,}\cite{Gyulassy:2004zy}. 
On the other hand, at lower energies, at the AGS, the maximum energy density is 
probably not sufficient to produce such a state of matter. 
Therefore, an energy scan progam in the intermediate energy range, covered by the CERN SPS, was initiated. 
The goal was to find anomalies in the energy dependence which could be related to the change of the number of degrees of freedom 
connected to a phase transition of the hot and dense matter created in heavy ion collisions.

In the course of the SPS energy scan program, the NA49 collaboration measured the hadronic final state of central Pb+Pb collisions 
at 20, 30, 40, 80, and 158 AGeV beam energy. In the following, the energy dependence of selected hadronic observables will be discussed,
with the main focus on where we start to see indications of a deconfined phase of matter.

\section{Experiment}
The NA49 detector system\cite{Afanasev:1999iu} is a large acceptance fixed-target hadron spectrometer. The main devices are four large TPC's. Two of them are placed
inside two super-conducting dipole magnets. The two other are installed downstream of the magnets left and right of the beam line. 

The momentum of a particle is determined by measuring its deflection in the magnetic field. The accuracy of the momentum reconstruction
is of the order of $\Delta p/p^2 \approx \mbox{0.3--7} \times 10^{-4}$(GeV/$c$)$^{-1}$.

A measurement of the specific energy loss (dE/dx) of a particle traversing the detector gas  
allows for particle identification. However, this method is only applicable if the particle momentum is larger than 4 GeV/c. Therefore,
 the NA49 detector is also equipped with two time of flight walls, which provide excellent particle identification in a small acceptance window
at midrapidity.

The centrality selection is based on a measurement of the energy deposited by the projectile spectator nucleons in a forward calorimeter.

The data presented here were taken between 1996 and 2002. 
For the samples at 20, 30, 40, and 80 AGeV the online trigger selected the most central 7\% of the total inelastic Pb+Pb cross section. In total,
roughly 1.5M events were recorded at these energies. At 158 AGeV, two data samples with 0.8M and 3M events were collected, at 5\% and 23.5\% centrality
respectively.

\section{Energy Dependence of Transverse Mass Spectra}
Transverse momentum spectra at midrapidity are shown in Figure 1. In the upper row preliminary results
at 20 and 30 AGeV are plotted, on the lower row spectra at 158 AGeV are shown.  

Assuming a common transverse flow velocity and a common thermal temperature for all different particles types,
a 'blast wave' parametrization\cite{Schnedermann:1993ws} was fitted to the data. It describes the data reasonably well
and we find at SPS energies for the freeze out temperature $T_f \approx 120 MeV$ and for the transverse velocity ${\beta}_T \approx 0.5$. 
These parameters do not change significantly in the SPS energy range.

\begin{figure}[h]
\begin{center}
\includegraphics[width=4cm]{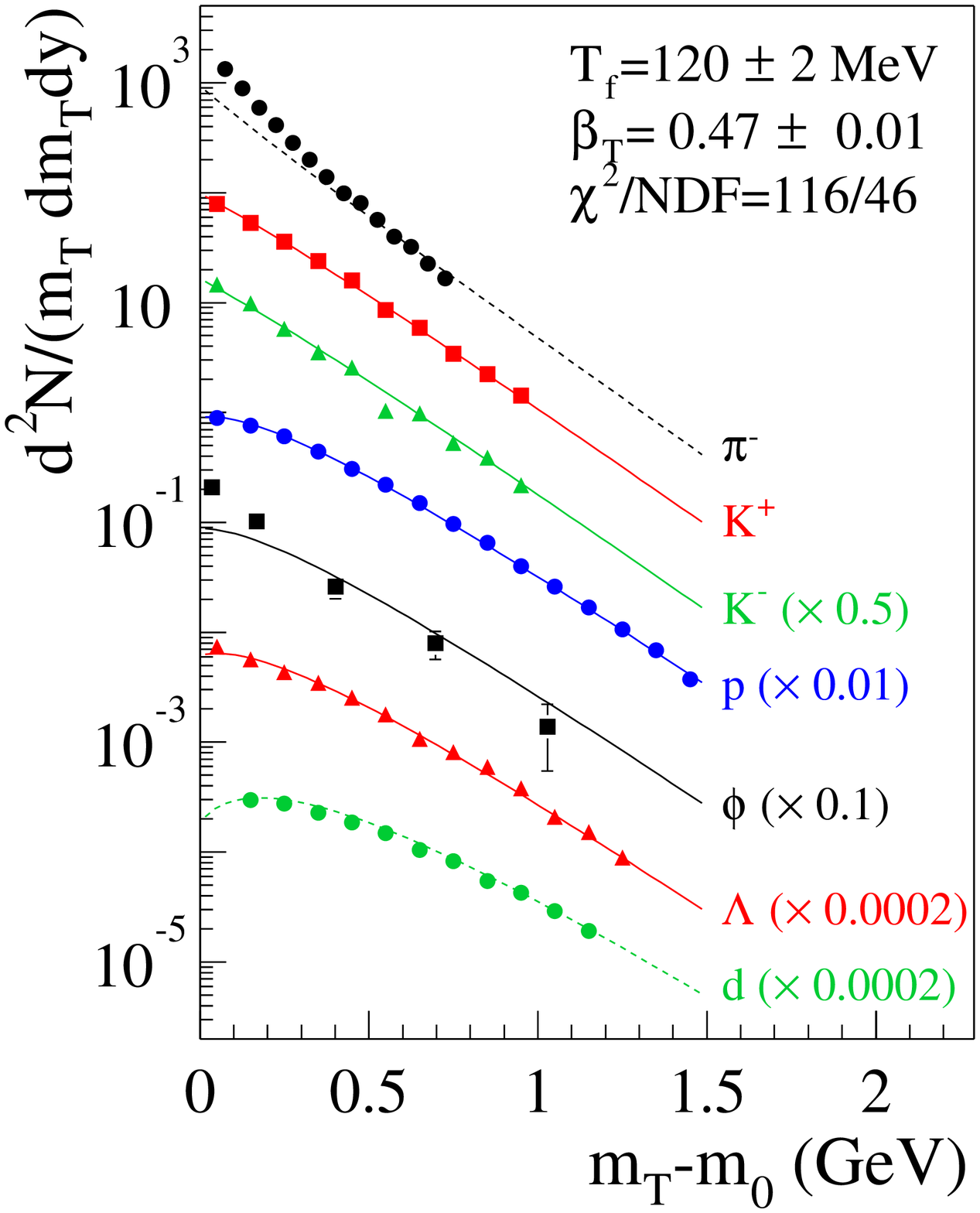}
\includegraphics[width=4cm]{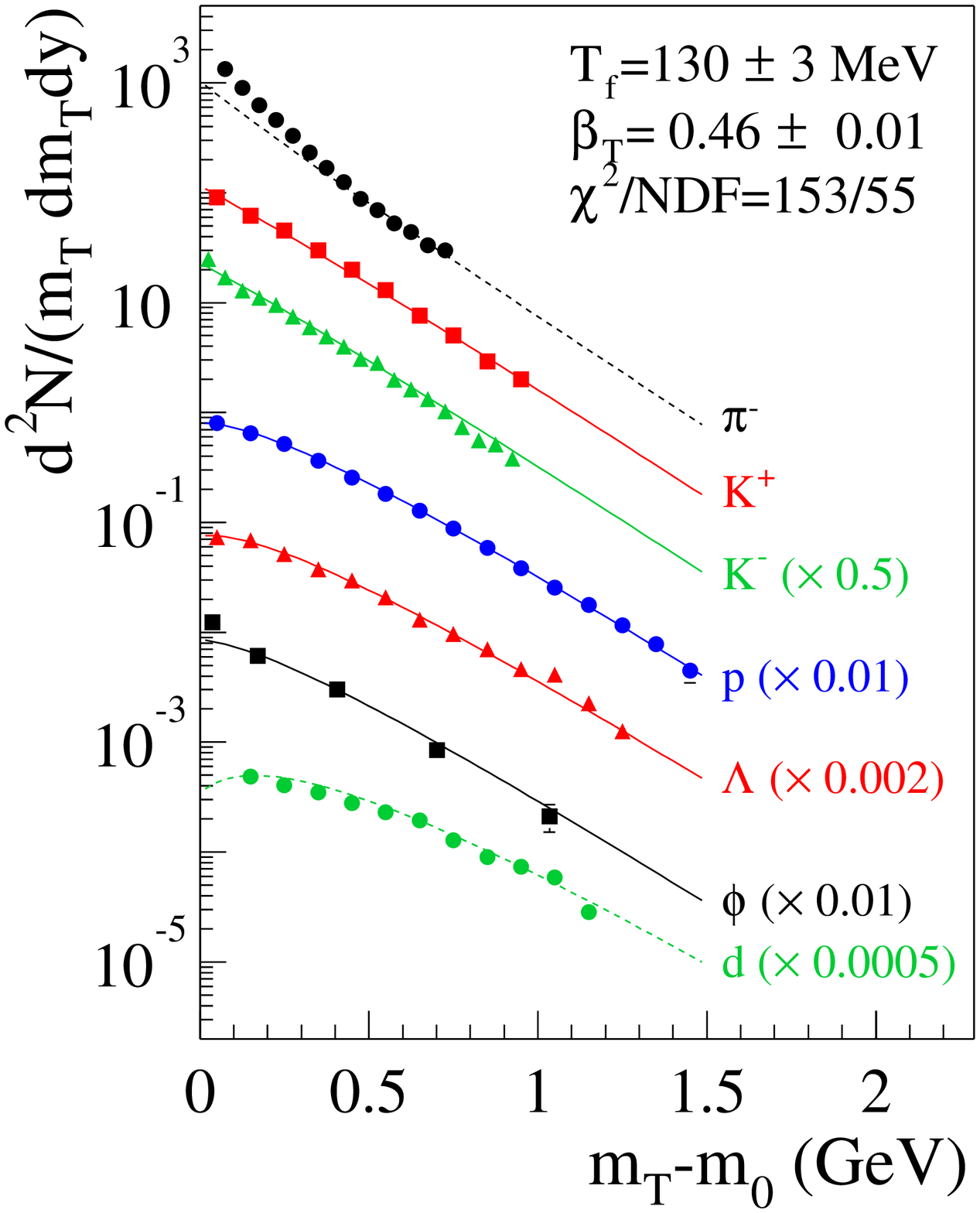}
\includegraphics[width=6cm]{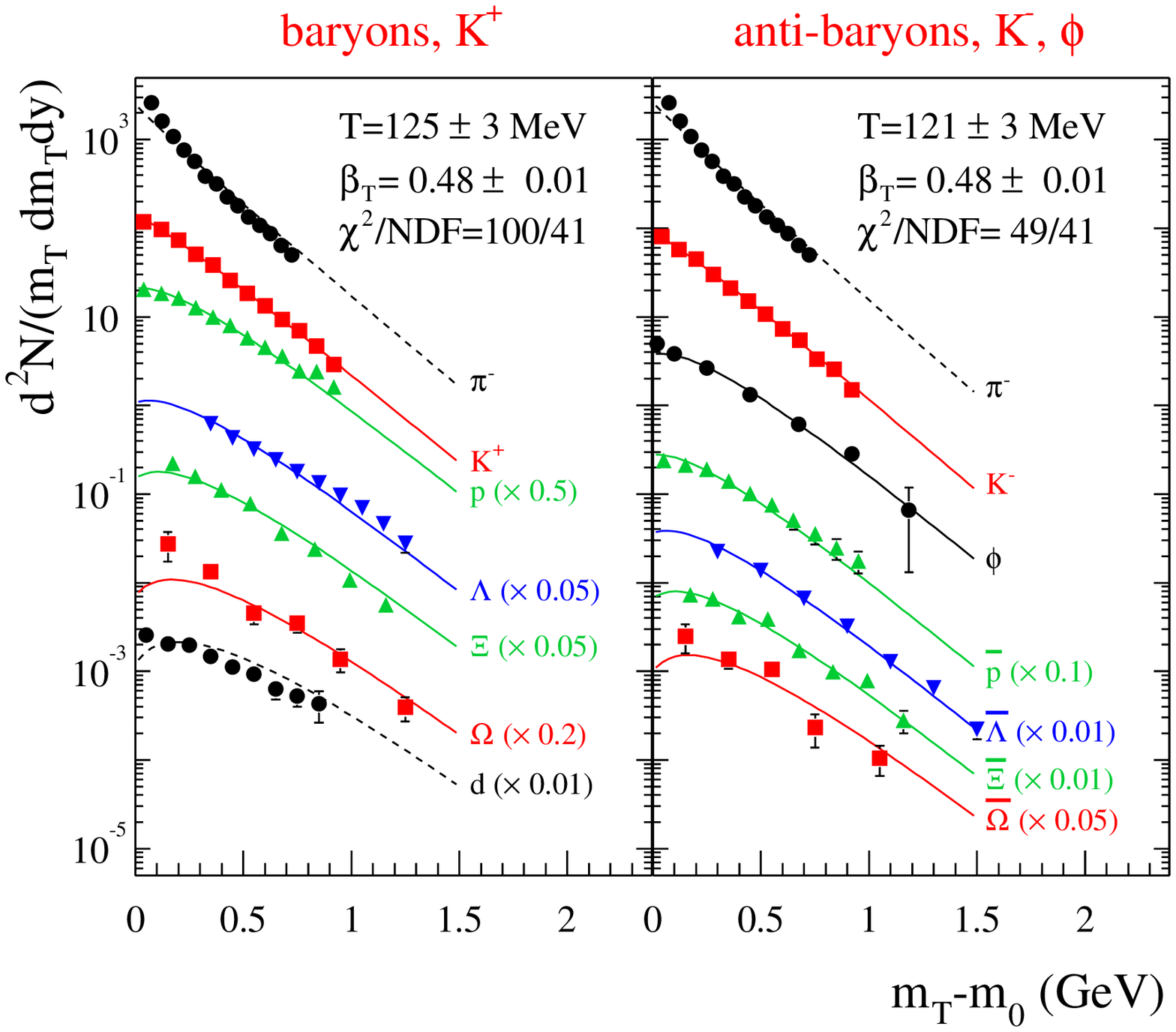}
\caption
{
  Transverse mass spectra at midrapidity.
  Lines and parameters $T_f$ and ${\beta}_T$ were obtained from blast wave fits. 
  \newline
  Upper row: preliminary results at 20 and 30 AGeV.
  Lower row: published results at 158 AGeV.
}
\label{fig:blastwave}
\end{center}
\end{figure}

A model independent way to compare transverse momentum spectra, is to consider the mean transverse mass 
$\langle{m_t}\rangle - m_0$.
In Figure 2 the energy dependence of this quantity is shown for pions, kaons and protons.

If energy density and hence pressure increase with beam energy, a stronger transverse expansion is
expected at higher energies.  Assuming that the strength of the transverse expansion is reflected in the mean transverse mass, 
the observable $\langle{m_t}\rangle - m_0$ will rise with $\sqrt{s_{NN}}$.

\begin{figure}[h]
\begin{center}
\includegraphics[width=8cm]{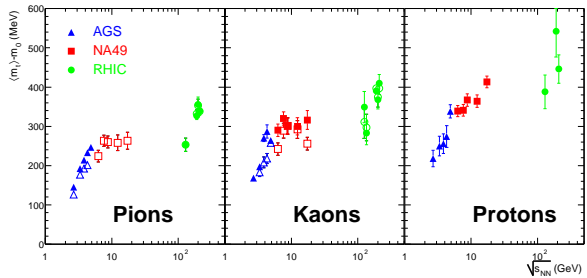}
\caption{Energy dependence of the mean transverse mass of pions, kaons and protons. Open symbols indicate negatively charged particles.}
\label{fig:engdep}
\end{center}
\end{figure}

At AGS energies we indeed observe a strong increase of the mean transverse mass of pions, kaons and protons with energy. However, at lower SPS energies this behaviour
changes: the mean transverse mass increases only weakly with beam energy.

This charateristic change in the energy dependence has been interpreted as a signature for a phase transition\cite{Gazdzicki:1998vd}. Because
the equation of state - the relation between energy density and pressure - changes at the phase boundary and with it the strength of the transverse expansion.

This observation supports the hypothesis, that the initial energy density reached in central A+A collisions at lower SPS energies is already large enough, 
to force a phase transition of the strongly interacting nuclear matter.

\section{Energy Dependence of Particle Yields}
The NA49 apparatus features a large acceptance in the forward hemisphere allowing for measurements of rapidity spectra from midrapidity up to almost beam rapidity.
The rapidity spectra for $\pi^-$, $\rm K^+$, $\rm K^-$, $\phi$, $\Lambda$ and $\overline \Lambda$ at five energies are shown in Figure 3. 
The spectra of the mesons can be parametrized by
Gaussians or double Gaussians. From these parametrizations the total particle yields are derived.
The rapidity distribution of $\Lambda$ particles looks different, since the production of a $\Lambda$ is more sensitive
to the baryon number distribution.

\begin{figure}[h]
\begin{center}
\includegraphics[width=14cm]{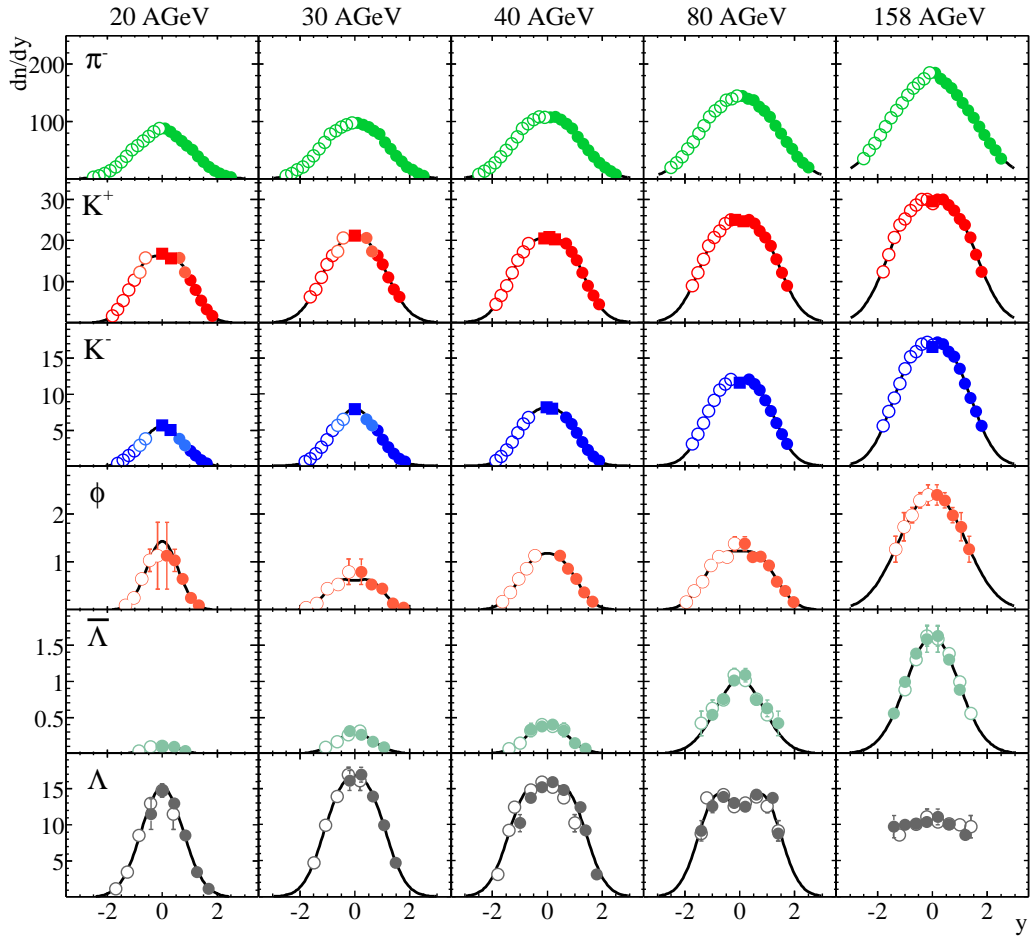}
\caption{Rapidity spectra of $\pi^-$, $\rm K^+$, $\rm K^-$, $\phi$, $\Lambda$ and $\overline \Lambda$ at five energies.
Lines indicate Gaussian or double Gaussian fits. Open symbols indicate data points reflected around midrapidity.}
\label{fig:rapidity}
\end{center}
\end{figure}

The total pion multiplicity per wounded nucleon as a function of Fermis measure $F \approx \sqrt{\sqrt{s_{NN}}}$ is shown in Figure 4. 
The relative pion production increases with collision energy, but starting from lower SPS energies the rate of increase grows. 
This feature is not visible in p+p collisions. 
The steepening of the energy dependence can be explained by an increase of the effective number of degrees of freedom due to the onset of deconfinement at
about 30 AGeV beam energy\cite{Gazdzicki:1998vd}.

\begin{figure}[h]
\begin{center}
\includegraphics[width=5.5cm]{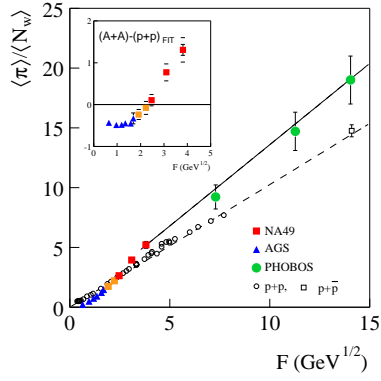}
\caption{Energy dependence of the mean number of pions per wounded nucleon in A+A and p+p collisions. Lines are only drawn to guide the eye.}
\label{fig:allpart}
\end{center}
\end{figure}

Figure 5 shows the energy dependence of relative strangeness production. The ratios of the total multiplicities ${\langle \rm K^+ \rangle} / {\langle\pi^+\rangle}$, 
${\langle \rm K^- \rangle} / {\langle\pi^-\rangle}$, 
${\langle \Lambda \rangle} / {\langle\pi\rangle}$  and ${\langle \overline \Lambda \rangle} / {\langle\pi\rangle}$  are compared to model calculations and when available to p+p collisions. 
While  ${\langle \rm K^- \rangle} / {\langle\pi^-\rangle}$ and ${\langle \overline \Lambda \rangle} / {\langle\pi\rangle}$ ratios rise continuously with energy,
 a distinct maximum is visible in the energy dependence of the ratios ${\langle \rm K^+ \rangle} / {\langle\pi^+\rangle}$ and ${\langle \Lambda \rangle} / {\langle\pi\rangle}$.
Hadron gas\cite{Cleymans:1999st} (HGM) and microscopic models\cite{Weber:2002pk} (RQMD, UrQMD) roughly describe the trend of the energy dependence, but especially the pronounced peak in 
the ${\langle \rm K^+ \rangle} / {\langle\pi^+\rangle}$ ratio is not reproduced. 
 As demonstrated in Figure 5, this feature is also absent in p+p collisions.    

\begin{figure}[h]
\begin{center}
\includegraphics[width=15cm]{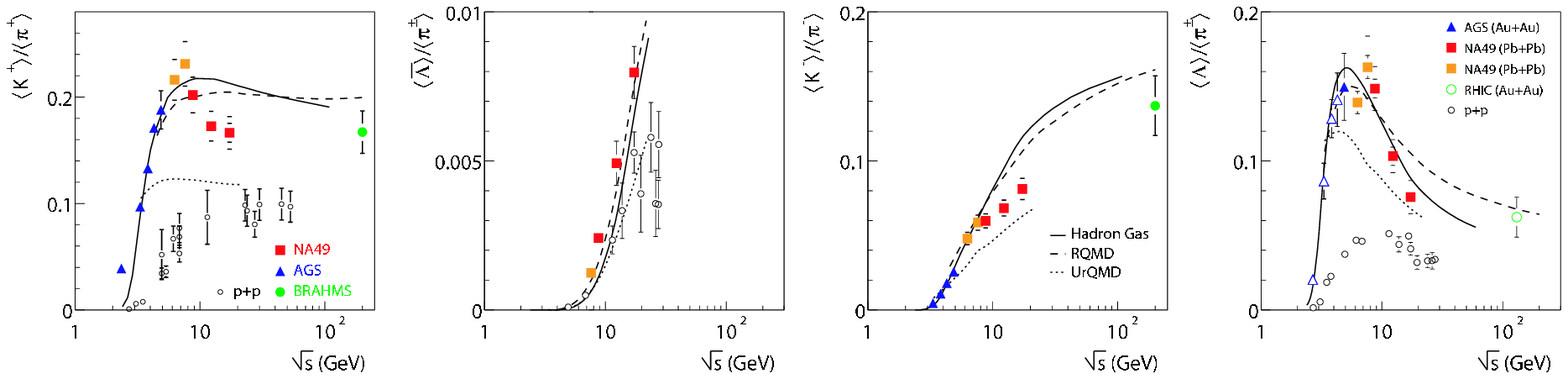}
\caption{Energy dependence of particle ratios  ${\langle \rm K^+ \rangle} / {\langle\pi^+\rangle}$, 
${\langle \rm K^- \rangle} / {\langle\pi^-\rangle}$, 
${\langle \Lambda \rangle} / {\langle\pi\rangle}$  and ${\langle \overline \Lambda \rangle} / {\langle\pi\rangle}$ in central A+A collisions and p+p collision.
Lines indicate model calculations.}
\label{fig:strangeness}
\end{center}
\end{figure}

\newpage

The total strangeness production to pion rate can be approximated by $E_S = (2(\langle \rm K^- \rangle + \langle \rm K^- \rangle) + \langle \Lambda \rangle) / \langle \pi \rangle$.
As seen from Figure 6, we find that the hadron gas and microscopic models describe the trend of the data, but they do not show the 
distinct maximum at lower SPS energies. Only a model\cite{Gazdzicki:1998vd} (SMES) including a phase transition around 30 AGeV  exhibits a peak as seen in the data.

\begin{figure}[h]
\begin{center}
\includegraphics[width=5.5cm]{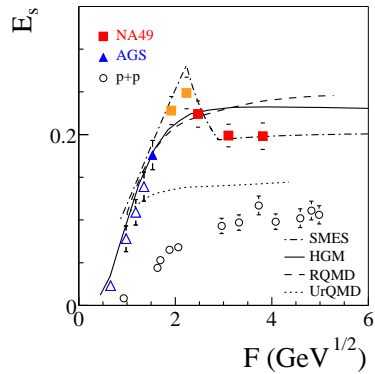}
\caption{Energy dependence of the strangeness to pion ratio and model calculations.}
\label{fig:es}
\end{center}
\end{figure}

\section{Summary and Outlook}
The energy dependence of several observables show anomalies at lower SPS energies. These signatures indicate, that the energy density reached in A+A collisions
at beam energies of about 30 AGeV are already sufficient to produce a plasma phase.

The next question to address is, how the system behaves when the volume is varied. At which intermediate system size appear indications for a deconfined phase? 
Future experiments at the SPS could deliver the answer to this question if the proposed light ion program will be approved\cite{Letter:2003aa}.

\section*{Acknowledgments}
This work was supported by the US Department of Energy
Grant DE-FG03-97ER41020/A000,
the Bundesministerium fur Bildung und Forschung, Germany, 
the Polish State Committee for Scientific Research (2 P03B 130 23, SPB/CERN/P-03/Dz 446/2002-2004, 2 P03B 04123), 
the Hungarian Scientific Research Foundation (T032648, T032293, T043514),
the Hungarian National Science Foundation, OTKA, (F034707),
the Polish-German Foundation, and the Korea Research Foundation Grant (KRF-2003-070-C00015).

\end{document}